# Information Cocoons in Online Navigation

Lei Hou[1,2], Xue Pan[1,2], Kecheng Liu[2], Zimo Yang[3], Jianguo Liu[4*], Tao Zhou[5*]

[1]School of Management Science and Engineering, Nanjing University of Information Science and Technology, Nanjing 210044, China.

[2]Informatics Research Centre, Henley Business School, University of Reading, Reading RG6 6UD, United Kingdom.

[3]Beijing AiQiYi Science & Technology Co. Ltd., Beijing 100080, China.

[4]Institute of Accounting and Finance, Shanghai University of Finance and Economics, Shanghai 200433, China

[5]Big Data Research Center, University of Electronic Science and Technology of China, Chengdu 611731, China.

*Correspondence to: liujg004@ustc.edu (J. Liu); zhutou@ustc.edu (T. Zhou).

**Abstract:** Social media and online navigation bring us enjoyable experience in accessing information, and simultaneously create information cocoons (ICs) in which we are unconsciously trapped with limited and biased information. We provide a formal definition of IC in the scenario of online navigation. Subsequently, by analyzing real recommendation networks extracted from *Science*, *PNAS* and *Amazon* websites, and testing mainstream algorithms in disparate recommender systems, we demonstrate that similarity-based recommendation techniques result in ICs, which suppress the system navigability by hundreds of times. We further propose a flexible recommendation strategy that solves the IC-induced problem and improves retrieval accuracy in navigation, demonstrated by simulations on real data and online experiments on the largest video website in China.

The explosive development of information techniques and services, in particular the emergence of portal sites, recommender systems, search engines and social media, has led us to a world of abundant information. We access diverse information via increasing sources, yet it is widely believed that information cocoons (ICs) are very often created in which we are unconsciously trapped with limited and biased information (1). The proliferation of ICs may result in an increase in social fragmentation, polarization and extremism, and eventually intensify segregation and threaten democracy (1-3).

Contributing factors to ICs are various, which can be roughly classified into two categories. (i) Active selection: individuals tend to link to like-minded people, read and produce articles with similar opinions, and ignore different voices, hence forming echo chambers (4-7). (ii) Passive choice: search engines and recommender systems feed information to users according to their past records, creating filter bubbles that narrow their navigation scopes (8-10). In addition, the above two types of behavior may coact and reinforce ICs via friend recommendation and news recommendation (11-13).

Although IC-related issues are under the spotlight of investigation and heated debates (4,14-17), quantitative studies about the existence and influence of ICs are rare, largely due to the lack of an explicit definition of IC and thus a benchmark for quantitative analyses. This paper will provide a mathematically formal definition of IC in a common scenario of online navigation, namely the recommendation network (RN) that connects similar contents with hyperlinks according to algorithmic evaluations (18-20). Denoting $G(V, E)$ a directed recommendation network where $V$ and $E$ are a set of nodes (objects) and a set of directed links (hyperlinks), then an IC is defined as a subset $C \subseteq G$ such that (i) the subgraph $G[C]$ induced by $C$ is strongly connected, and (ii) there is no outgoing link from a node in $C$ to a node outside $C$. A node belonging to an IC is called as IC node (note, a node at most belongs to one IC) and the number of nodes in an IC is called its size.

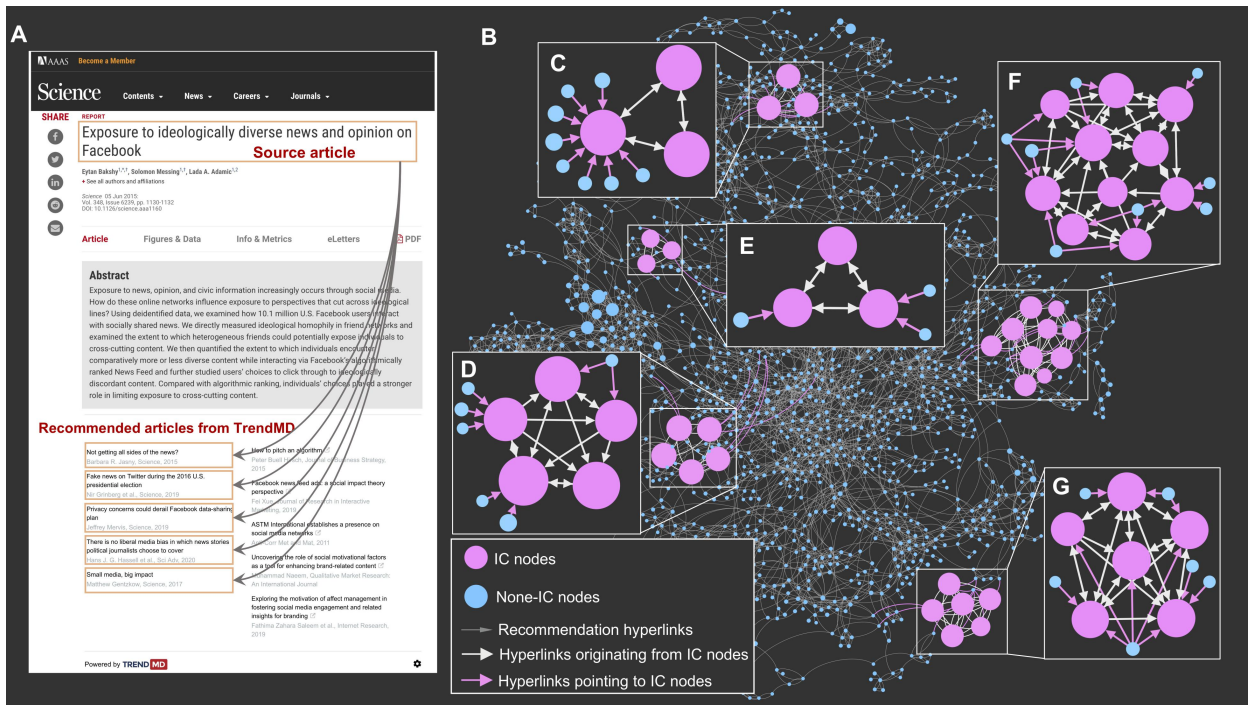

**Fig. 1.** The recommendation network (RN) of Science articles. (**A**) Screenshot of an article's webpage from Science, where a list of recommended articles is displayed with hyperlinks embedded. (**B**) A sample of collected Science RN, where each node is an article, and each directed link is a hyperlink. The node sizes are proportional to the logarithm of visiting frequencies of a random walk. (**C-G**) Showcases of five information cocoons and their neighboring nodes, which are empirically observed in the Science RN.

We firstly examine three empirical RNs extracted from websites of *Science*, *PNAS* and *Amazon* (see Supplementary Text S1 for detailed description). Figure 1 illustrates the case of *Science*. In the website of *Science*, the page of each research article lists several recommended articles with hyperlinks embedded (Fig. 1A). These hyperlinks constitute the *Science* RN (Fig. 1B), aiming at helping users explore relevant information. Table 1 presents fundamental statistics of the three empirical RNs. According to the definition of IC, there are 96, 79 and 1181 ICs for *Science*, *PNAS* and *Amazon* RNs, respectively (Supplementary Text S2 introduces the algorithm to identify ICs, and Fig. 1C-1G show five typical ICs in *Science* RN).

**Table 1.** Statistics for the studied recommendation networks, where the symbol # stands for the number of and IC traffic means the percentage of visits on IC nodes during an *N*-step random walk.

| RN | #Objects | #ICs | #IC nodes | IC traffic | Navigability |
|---|---|---|---|---|---|
| Empirical RNs | | | | | |
| Science | 7,730 | 96 | 350 | 94.77% | 0.44% |
| PNAS | 59,479 | 79 | 415 | 96.98% | 0.67% |
| Amazon | 119,636 | 1,181 | 10,859 | 95.81% | 0.07% |
| Derived RNs | | | | | |
| Steam | 10,978 | 10 | 88 | 99.98% | 0.07% |
| Yelp | 60,785 | 8 | 135 | 16.14% | 0.09% |
| Epinions | 61,273 | 3 | 41 | 99.98% | 0.05% |
| MovieLens | 33,670 | 3 | 20 | 99.99% | 0.03% |

We apply random walks (21) to simulate users' surfing activities. Given an RN with $N$ nodes, its navigability $\Omega(G)$ can be defined as the expected coverage of distinct nodes being visited during an $N$-steps random walk from a randomly selected starting node (22). Denoting $n(t)$ the expected number of distinct nodes being visited during a $t$-steps random walk, for a completely random network with same existence likelihood of every potential link, the growth of $n(t)$ follows the dynamics $\frac{d}{dt}n(t) = 1 - \frac{1}{N}n(t)$, namely $n(t) = N(1 - e^{-t/N})$. Hence the corresponding navigability is $\Omega = n(N)/N = 1 - 1/e \approx 63.2\%$. As shown in Table 1 and Fig. 2A, to our surprise, navigabilities of the three RNs are all less than 1%, while IC nodes monopolize most traffic (generally ≳ 95%, see Table 1). As the in-degree distribution of the three RNs are heavy-tailed (see Supplementary Text S1), it is also possible that bub nodes with large in-degrees dominate the traffic. To separate the effect from ICs and hub nodes, we apply the link-crossing operations sufficiently many times to get the corresponding first-order null network (23). In each operation, two links, say $a \to b$ and $c \to d$, are randomly selected and switched as $a \to d$ and $c \to b$. The selection ensures the avoidance of multiple link and loop. In a null network, degree sequence keeps unchanged while ICs are absent. As shown in Fig. 2A, in despite of the presence of hub nodes, $n(t)$ curves for null networks closely follow the prediction of random networks, suggesting that IC nodes rather than hub nodes result in the poor navigabilities. We have further demonstrated that a very few manipulated ICs, inserted into a completely random recommendation network, will lead to the above-observed poor navigability (see Supplementary Text S3).

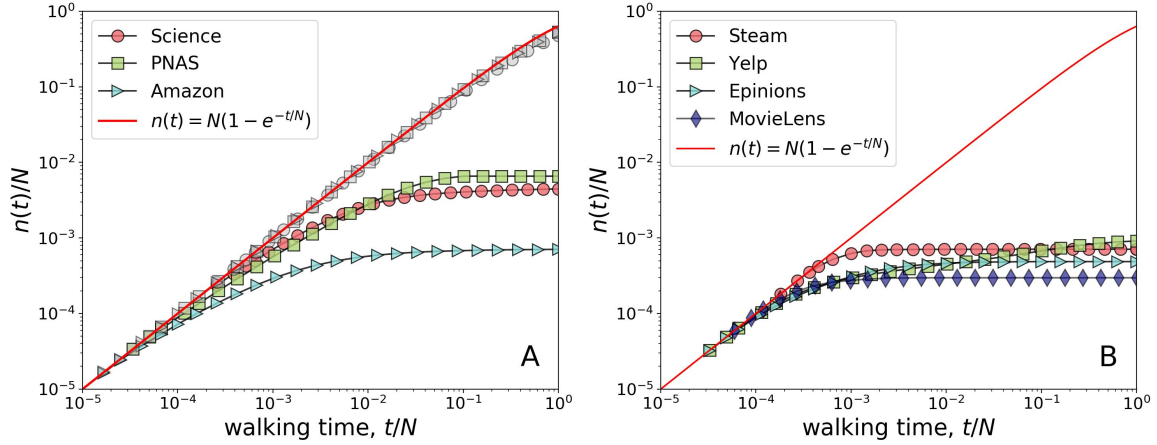

**Fig. 2.** Number of distinct nodes being visited during random walks in the (**A**) empirical RNs and (**B**) derived RNs for $L = 5$. The red curve in each plot denotes the prediction from a completely random network, and grey circles, squares and triangles in (**A**) represent the results for the null networks of *Science*, *PNAS* and *Amazon* RNs. For each network under consideration, the result is averaged over $N$ random walk experiments with every node being once the starting node.

Though it is very likely that links in empirical RNs are connecting similar objects, we do not know the exact mechanism underlying empirical RNs. Therefore, we next generate RNs by implementing mainstream recommendation algorithms based on real user-object interaction data sets. We consider four real data sets (*Steam*, *Yelp*, *Epinions* and *MovieLens*, see Supplementary Text S4 for details), each of which can be described by a bipartite network $G^B(U, O, E^B)$ where $U = \{u_1, u_2, \cdots, u_M\}$ is the set of users, $O = \{o_1, o_2, \cdots, o_N\}$ is the set of objects, and $E^B$ is the set of links between users and objects. According to widely-applied similarity-based recommendation techniques, a recommendation network $G$ can be generated by linking each object to its top-$L$ most similar objects with pairwise similarity being defined based on $G^B$. We adopt the common neighbor index (24) $s_{\alpha\beta} = \sum_{u \in U} b_{u\alpha} b_{u\beta} + \epsilon$, where $\epsilon \to 0$ is a tiny random number used to remove degeneracy caused by same similarity scores and $B^{M \times N}$ is the adjacency matrix of $G^B$ with $b_{uo} = 1$ if user $u$ connects with object $o$ and $b_{uo} = 0$ otherwise. Analogous to the empirical RNs, all four derived RNs have heavy-tailed in-degree distributions (see Supplementary Text S4), a few ICs that dominate the traffic (see Table 1), and much lower navigabilities in compared with random RNs (see Table 1 and Fig. 2B). The results for other well-known similarity indices are close (see Supplementary Text S4 for definitions and results of Jaccard index (24), Salton index (24) and heat conduction index (25)). In a word, the similarity-based recommendation algorithms can generate ICs and thus lead to poor navigability.

A possible cause for forming ICs is the reciprocity (if $\alpha$ is among the most similar objects to $\beta$, then $\beta$ is likely among the most similar objects to $\alpha$) and transitivity (if both $\alpha$ and $\beta$ are among the most similar objects to $\gamma$, then $\alpha$ and $\beta$ are likely to be very similar to each other) of similarity. This subsequently leads to local clusters if we simply pick up the top-$L$ most similar objects to construct the RN, and ICs are just extreme clusters. To break ICs and thus improve navigability, we suggest a flexible recommendation strategy that selects the $L$ outgoing links of each object from its top-$\lambda L$ ($\lambda \geq 1$) most similar objects (see Fig. 3A for an illustration). As shown in Supplementary Text S5 and Fig. 3B, the increasing $\lambda$ quickly reduces ICs and largely improves navigability. Meanwhile, we should also consider the effect of $\lambda$ on the ability to hit a user's interests. To quantify such ability, in each user-object interaction data set, users are randomly divided into a training group and a testing group, and only the information of training users is used to construct the RN. Each testing user $u$ then performs a random walk starting from one of $u$'s selected objects. After $t$ time steps, the hitting rate of $u$'s interests is $r_u(t) =$ $h_u(t)/(k_u-1)$, where $k_u$ is the number of selected objects of $u$ (i.e., the degree of $u$ in the original user-object bipartite network) and $h_u(t)$ is the number of visited objects among the $k_u - 1$ selected objects (except for the starting object) during the $t$-step random walk. The overall retrieval accuracy $r(t)$ is the average of hitting rates over all testing users. As shown in Fig. 3C, the optimal value of $\lambda$ subject to the largest $r(t)$ is larger than 1 unless $t$ is very small, and there exists a huge area in the $(\lambda, t)$ plane wherein navigability and retrieval accuracy can be simultaneously improved.

We further test whether the flexible recommendation strategy is effective in a real scenario of online navigation. The experiment was implemented in AiQiYi (NASDAQ: IQ), the largest video website in China with about $1.5 \times 10^8$ daily active users and $5 \times 10^8$ monthly active users. To fill a recommendation position, relevant videos are selected by a series of recall algorithms from all candidates and then sorted by a ranking model, the top item that can pass the final regulation (to filter out violent, porno and brand-conflicting videos) will be exhibited (see Supplementary Text S6 for an illustration of the structure of AiQiYi's recommender system). The item-based collaborative filtering (ICF) is one of recall algorithms, which finds out the most relevant videos according to recently clicked videos of the target user. Under each request, the original ICF returns the top-5 most relevant videos, and in our experiment, for users in the treatment group, it returns 5 randomly selected videos from the top-10 most relevant ones, corresponding to the flexible recommendation strategy with $\lambda = 2$. The experiment was deployed in the first two positions of the *Guess You Like* column in the landing page, which are hottest positions attracting about $1.5 \times 10^8$ clicks from about $8 \times 10^7$ distinct users per day. To evaluate the performance, we employ two widely used metrics in industry, playing rate (PR) and playing duration (PD). The former is the ratio of playing to clicking of recommended videos, and the latter is the average playing duration. The precise definitions and the rationale of the choice of these two metrics are shown in Supplementary Text S6. The experiment lasted one week from Nov. 3 to Nov. 9 in 2020 (daily results are presented in Supplementary Text S6), with average PR and PD over seven days being 73.22% and 61.37 minutes for the treatment group (5% users, randomly selected), and 73.12% and 61.33 minutes for the control group (95% users). The experiment only changes a tiny part of an elaborately-designed and well-trained recommender system in real industry but bring about 150,000 more video plays per day (the change of PR is statistically significant, see $t$-test in Supplementary Text S6), again indicating the effectiveness of the flexible recommendation strategy.

In despite of ongoing and heated debates on the harmfulness of ICs (4,14-17), a formal definition of IC is lacking. The primary contribution of this paper is to provide a mathematically explicit definition of IC, and to demonstrate the existence and notably negative effect on navigability of ICs in both empirical and derived recommendation networks. The definition seems too strict and thus less applicable, however, based on its core idea, it can be extended to characterize more substructures of directed networks. For example, the extent a strongly connected subgraph $G[C]$ induced by a node set $C$ is likely to be an IC can be measured by its escaping probability $p_e(C)$, defined as the ratio of escaping links (links from nodes in $C$ to nodes outside $C$) to total links starting from nodes in $C$. Then, a strongly connected subgraph with $p_e$ no more than a preseted threshold can be treated as a quasi-IC (QIC). Such extension energizes our analyses, for example, as shown in Supplementary Text S7, the two QICs in Yelp, respectively of escaping probabilities 0.0111 and 0.0118, dominate 71.86% of the random walk traffic, well explained the remarkably lower IC traffic of Yelp in Table 1.

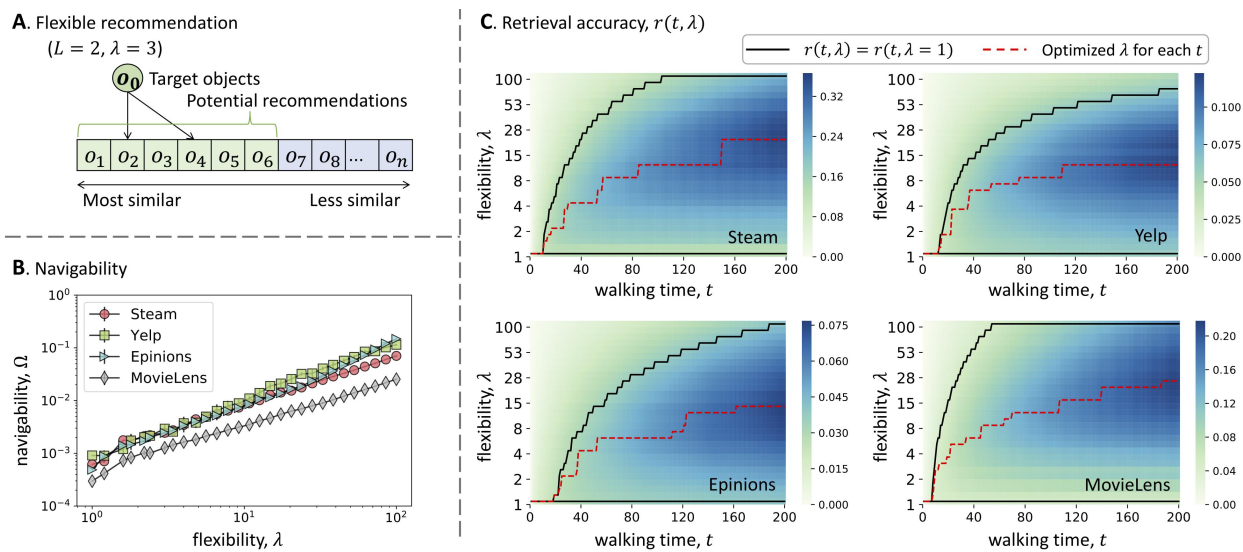

**Fig. 3.** Efficacy of the flexible recommendation strategy. (**A**) Illustration of the flexible recommendation strategy, where the target object randomly recommends $L$ objects from the pool of $\lambda L$ most similar ones. (**B**) Navigability of the four derived RNs with different $\lambda$. (**C**) Heatmaps for retrieval accuracy in the $(\lambda, t)$ plane, where the black solid curves mark the areas with improved accuracy. The red dashed lines indicate the optimal $\lambda$ subject to the highest accuracy. The results reported in (**B**) and (**C**) are obtained based on the common neighbor index and a 90%-10% division of the training and testing groups. For each data set, given the flexibility $\lambda$, the navigability is averaged over 100 realizations of RNs, and for each RN, the result is averaged over $N$ random walk experiments with every node being once the starting node. The retrieval accuracy is also averaged over 100 realizations of RNs, and for each RN, 5 independent random walk experiments are implemented for each pair of testing user and possible starting object.

Similarity-based recommendation algorithms used to be popular, and are still important modules in industrial recommender systems up to date (18,26). Present simulations on similarity-based algorithms thus indicate that recommender systems, by nature of their design, tend to insulate users from exposure to diverse content. Recent ethical studies (10,27,28) have noticed this issue and suggested the avoidance of filter bubbles as a high-priority task in designing or improving recommender systems, but they do not provide a viable pathway towards the target. Nearest acts and regulations emphasize algorithmic transparency (see *Algorithmic Justice and Online Platform Transparency Act*, as a bill in the Senate of the United States) and users' rights in shutting down recommending services and deleting part or all of personalized tags (see *Management Regulations on Algorithmic Recommendations in Internet Information Services*, as a exposure draft by the Cyberspace Administration of China). However, these rules cannot pull us out of ICs because filter bubbles are not resulted from opacity and users are usually not aware of (or even enjoying) biased information. What's worse is that such arts and regulations, if not being rightly applied, may reduce out benefits from algorithms underlying online navigation. Different from known ethical suggestions and legal rules, our results indicate that IC-related problems can be largely solved inside the algorithmic framework, using the proposed flexible strategy or other alternatives (8,29,30). This paper utilizes mathematical concepts, quantitative analyses and computational tools to describe, characterized and solve IC-related problems, which is also referential to other problems in technical ethics.

# References

1. C. R. Sunstein, Ed., *Infotopia: How many minds produce knowledge.* (Oxford University Press, 2006).

2. N. J. Stroud, Polarization and partisan selective exposure. *Journal of Communication* **60**, 556-576 (2010)

3. C. R. Sunstein, Is social media good or bad for democracy. *International Journal of Human Rights* **27**, 83-89 (2018)

4. E. Bakshy, S. Messing, L. A. Adamic, Exposure to ideologically diverse news and opinion on Facebook. *Science* **348**, 1130-1132 (2015).

5. J. Wihbeym K. Joseph, D. Lazer, The social silos of journalism? Twitter, news media and partisan segregation. *New Media & Society* **21**, 815-835 (2019).

6. J. Hu, Q.-M. Zhang, T. Zhou, Segregation in religion networks. *EPJ Data Science* **8**, 6 (2019).

7. M. Mosleh, C. Martel, D. Eckles, D. G. Rand, Shared partisanship dramatically increases social tie formation in a Twitter field experiment. *Proc. Natl. Acad. Sci. U.S.A.* **118**, e2022761118 (2021).

8. T. Zhou *et al.*, Solving the apparent diversity-accuracy dilemma of recommender systems. *Proc. Natl. Acad. Sci. U.S.A.* **107**, 4511-4515 (2010).

9. E. Pariser, *The filter bubble: What the Internet is hiding from you* (Penguin, London, UK, 2011).
12